# Tabletop single-shot extreme ultraviolet Fourier transform holography of an extended object

**Erik B. Malm,[1,*] Nils C. Monserud,[1] Christopher G. Brown,[1] Przemyslaw W. Wachulak,[2] Huiwen Xu,[3] Ganesh Balakrishnan,[3] Weilun Chao[4], Erik Anderson[4] and Mario C. Marconi[1]**

[1]Engineering Research Center for Extreme Ultraviolet Science and Technology, and Electrical and Computer Engineering Department, Colorado State University, Fort Collins, Colorado 80523, USA
[2]Institute of Optoelectronics, Military University of Technology, ul. gen. S. Kaliskiego 2, 00-908 Warsaw, Poland
[3]Center for High Technology Materials, and Department of Electrical and Computer Engineering, University of New Mexico, Albuquerque, New Mexico 87106, USA
[4]Center for X-Ray Optics. Lawrence Berkeley National Lab, 1 Cyclotron Rd, Berkeley, CA 94720, USA
*malm@rams.colostate.edu

**Abstract:** We demonstrate single and multi-shot Fourier transform holography with the use of a tabletop extreme ultraviolet laser. The reference wave was produced by a Fresnel zone plate with a central opening that allowed the incident beam to illuminate the sample directly. The high reference wave intensity allows for larger objects to be imaged compared to mask-based lensless Fourier transform holography techniques. We obtain a spatial resolution of 169 nm from a single laser pulse and a resolution of 128 nm from an accumulation of 20 laser pulses for an object ~11x11μm$^2$ in size. This experiment utilized a tabletop extreme ultraviolet laser that produces a highly coherent ~1.2 ns laser pulse at 46.9 nm wavelength.

## References and links

## 1. Introduction

Extreme ultraviolet (EUV) and soft X-ray microscopy has shown the capability for nanometer spatial and femtosecond temporal resolutions and promises to help explore nanoscale dynamics in the physical and life sciences. Lensless Fourier transform holography (FTH) and iterative phase retrieval have emerged as leading short wavelength coherent diffraction imaging techniques. FTH and iterative phase retrieval are complementary techniques. Mask-based FTH has a more complicated mask fabrication procedure, but the object reconstruction process is fast and simple. The resolution is limited by the reference features fabricated within the mask. In contrast, the resolution in iterative phase retrieval is unrestricted by the fabrication of the mask or optics, but has a longer and more complicated reconstruction procedure, that presently, does not allow for real-time imaging.

Iterative phase retrieval recovers an object's exit field from its diffraction pattern [1,2]. *A priori* knowledge such as the object's size and positivity are often utilized to increase the convergence speed or ensure uniqueness for an otherwise underdetermined system. The ability to image a sample without the use of optics makes iterative phase retrieval well suited for EUV/X-ray high-resolution imaging experiments. This technique has been applied to: nanoscale imaging [3], pump-probe [4], polychromatic data [5,6], tomography [7,8], and has been used in combination with FTH [9,10]. Recently, iterative phase retrieval has been used for 3-D imaging from a single diffraction pattern [11]. Iterative phase retrieval has proven to be a useful and emerging technology in the X-ray coherent diffraction imaging field. Similar to iterative phase retrieval, FTH has shown broad application to X-ray microscopy with high-resolution capability. The demonstrated capabilities and advantages of coherent diffraction imaging made these techniques attractive alternatives for high-resolution imaging and has triggered several experiments in recent years [12,13].

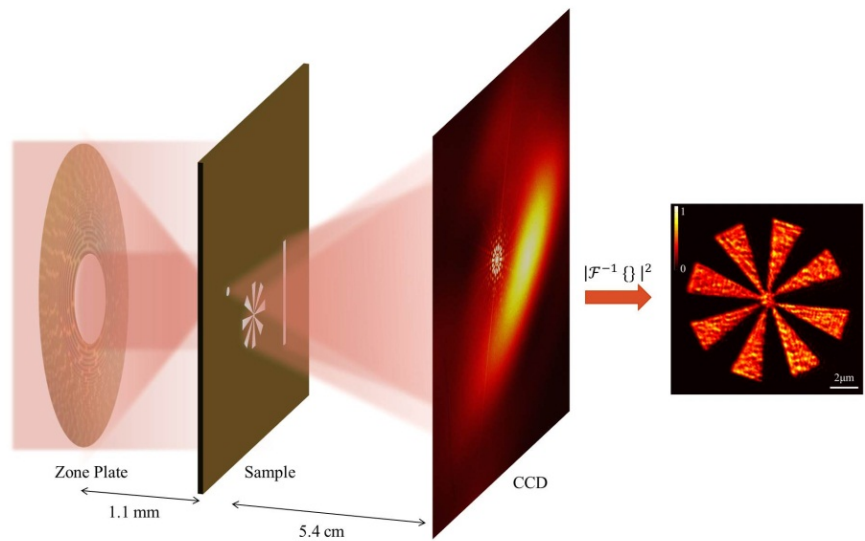

Fig. 1. Schematic of the Fourier transform holography setup. The 1$^{st}$ diffraction order of the Fresnel zone plate is used to create the reference wave. The central opening in the zone plate passes the incident beam directly to the sample. The reference and object waves interfere on the CCD and the object is numerically reconstructed by taking the modulus squared of the inverse Fourier transform of the interference pattern.

Fourier transform holography is a coherent imaging technique which utilizes the interference between reference and object waves to encode object information into the interference pattern [Fig. 1]. The object can be numerically reconstructed by applying the modulus squared to the inverse Fourier transform of the interference pattern collected on the detector. This simple reconstruction method allows for a fast retrieval of the object using the digitized hologram from a charge coupled device (CCD) detector. The rapid and deterministic reconstruction and high-resolution capability makes FTH an appealing imaging method. Conventional mask-based FTH uses a pinhole fabricated into the object mask to create the reference wave. Mask-based FTH and zone plate based holography have been used to image biological samples [14], magnetic nanostructures [15], for large field of view and sequential femtosecond imaging [16,17], and tomography [18,19]. For high-resolution reconstructions a small pinhole (or reference) is necessary; however, for a high quality reconstruction, the reference and object waves must be of similar intensities to optimize the fringe visibility [20]. Therefore, high-resolution mask-based FTH ultimately limits the object size that can be imaged. To overcome this drawback, multiple reference holes [21,22], ultra redundant arrays [23], and the development of holography with extended reference by autocorrelation linear differential operation (HERALDO) [24] have been utilized.

HERALDO is an extension of FTH which allows for larger reference apertures to be used while maintaining nanoscale resolution. It has been experimentally shown to provide high spatial resolution [25] and the capability for single-shot exposure with nanoscale resolution [26]. In contrast to FTH, the resolution of HERALDO is limited by the sharpness of the reference edges. Due to the reconstruction process, the high spatial frequencies in a hologram are amplified. The lower signal-to-noise ratio at the higher spatial frequencies coupled with the amplification from the application of the polynomial product can result in a degradation of the reconstruction. This makes HERALDO especially sensitive to noise.

In this paper, we present single-shot (1.2 ns temporal resolution) and multi-shot FTH with a tabletop EUV laser of an extended object. Our results demonstrate high-resolution images of an extended object ~11x11µm$^2$ in size. The high coherent flux of the tabletop EUV laser allowed for the recording of extended objects with 128 nm spatial resolution. The setup, utilized in this paper, is similar to the one used by Ian McNulty et. al. , in which the reference for FTH is created with a Fresnel zone plate [27].

## 2. Experimental setup

Figure 1 shows a diagram of the experimental setup utilized in this experiment. The first order focus of the zone plate was used to create the spherical reference wave. The central opening in the zone plate allowed for the incident beam to directly illuminate the object, which in this case, was a test pattern defined in a gold coated $Si_3N_4$ membrane. The membrane, where the object was defined, has a pinhole located next to the test pattern to be imaged. The pinhole selected the first diffraction order from the zone plate and filtered out the other orders. Due to the high reference intensity, this method overcomes the object size limitations in the mask-based FTH and HERALDO while maintaining nanoscale resolution.

In this setup, the spatial resolution is limited by the focal spot size of the zone plate. Unlike mask-based FTH, the pinhole size in this experiment only weakly affects the resolution of the reconstruction. The relationship between the pinhole radius and the resolution was determined numerically. Assuming a reference wave composed of a pinhole uniformly illuminated plus a focal spot in its center, we calculated the expected resolution of the holograms for different pinhole diameters and point source intensities. The effect of the pinhole size on the resolution depends upon the relative field strengths of the incident beam and the focal spot. As the focus intensity approaches zero, the resolution becomes highly dependent on the pinhole size and the relationship approaches that of the mask-based FTH. In contrast, as the focal spot intensity increases, the resolution becomes less dependent on the pinhole size. In our setup, the focus is very intense and therefore the resolution is weakly affected by the pinhole size. However, it is important to use a small pinhole to reduce any effects from the pinhole field, but more importantly the small pinhole reduces the effect of having a central opening in the zone plate. This effect was experimentally verified by passing the first order zone plate focus through a larger pinhole in a similar FTH experiment. Using a larger pinhole resulted in a more uniform reference wave, but the reference wave lacked lower spatial frequencies which resulted in an edge enhanced reconstruction [28].

The coherent light pulse is generated by a tabletop EUV laser at 46.9 nm wavelength. A fast discharge creates a Ne-like $Ar^{+8}$ plasma within an alumina capillary [29]. The laser produces a pulse with a 1.2 ns time duration, 0.2 mJ of energy and 4 mrad beam divergence. The beam has a 750 µm coherence length and 550 µm coherence radius at the sample location [30,31]. The laser's high spatial and temporal coherence makes it well suited for coherent imaging applications. The plasma conditions and resulting index of refraction within the capillary create an annular beam profile [29]. After exiting the capillary, the beam reflects off a 45º Si/Sc multilayer mirror which provides additional spectral filtering, further discriminating the laser light from the plasma EUV background [32] and allows for precise alignment of the beam. The Si/Sc multilayer mirror was optimized for the laser wavelength (46.9 nm) with a peak reflectivity of 31%. At the zone plate position, the beam expanded to 20 mm in diameter and from this a small section (0.5 mm in diameter) was used to illuminate the zone plate. The large beam size, at the zone plate location, as compared with the zone plate diameter allowed for uniform illumination of the zone plate. After reflecting off the multilayer mirror, the beam impinged on the zone plate where it is split into the reference and object waves.

The zone plate that produced the reference wave and illuminated the object was mounted 2.3 m from the laser output. It was fabricated as a self-standing structure to maximize the transparency at the EUV laser wavelength. It has 1250 zones with an outer diameter of 500 µm and a 100 µm diameter central opening. The outer zone width is 100 nm which corresponds to a first order numerical aperture of 0.23. The zone plate focus was positioned inside the pinhole of the sample mask with an actuator driven 3-axis stage. The sample pinhole is 700 nm in diameter and located 29 µm from the object pattern.

The object was fabricated with focused ion beam milling of 200 nm $Si_3N_4$ layer coated with 200 nm of gold. The object was a Siemens star 11.5 µm in diameter. A slit 15 µm long was fabricated into the mask which was used to assist in locating the position of the pinhole during alignment. The center of the slit is located 29 μm from the center of the Siemens star. To match the object and reference intensities, previous experiments typically leave a silicon nitride layer that covers the sample pattern. Due to the high reference intensity in this setup, the pattern was milled completely through the sample. The diffraction pattern was collected 5.4 cm away from the sample mask with a back illuminated Peltier cooled Andor X-ray CCD camera with 13.5 µm square pixels and an overall size of 2048×2048 pixels in total. To limit intrinsic thermal CCD noise, the camera was cooled down to –20 ºC during image acquisition.

## 3. Results and discussion

To illustrate the capability to image an extended object, a hologram was acquired with an accumulation of 20-shots [Fig. 2(b)]. Interference fringes in the hologram are visible at the edges of the image indicating that the entire numerical aperture of the zone plate was utilized. The highest intensities in Fig. 2(b) and Fig. 3(a) were clipped for presentation purposes. A scanning electron micrograph (SEM) of the gold test object is shown in Fig. 2(a). The gray areas indicate the presence of a gold coating on a $Si_3N_4$ support and the black areas are empty space (through holes). Figure 2(c) shows the reconstruction obtained by taking the modulus squared of the inverse Fourier transform of the hologram. The reconstruction represents the intensity distribution of the EUV field in the object plane. The resolution limit is set by the focal spot size of the zone plate.

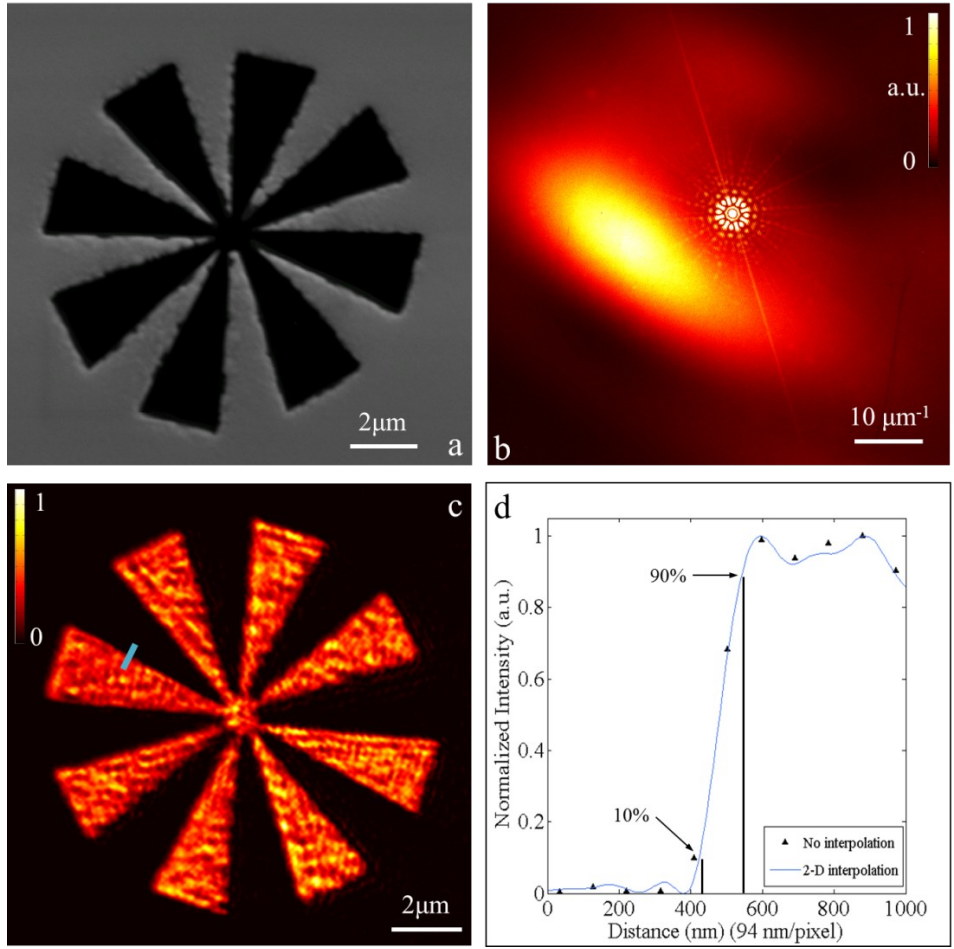

Fig. 2. (a) SEM of the Siemens star object which was fabricated with a focused ion beam. The gray areas indicate the gold layer and the black areas contain no material. (b) Hologram interference pattern from an accumulation of 20-shots collected on a CCD camera. (c) The normalized intensity reconstruction obtained from the hologram in (b). (d) A characteristic knife edge cut along the blue line in (c) with 120nm resolution. The blue line was obtained by zero padding the hologram. The discrete points were obtained without any zero padding of the hologram.

With this particular zone plate and mask, the theoretical resolution limit is 126 nm. To calculate the theoretical resolution, the first order focal spot size was calculated from the zone plate dimensions (inner and outer radii) and the outer zone width. The intensity outside of the pinhole was then set to zero and the intensity was normalized. It was converted into a line spread function and finally integrated to obtain the theoretical resolution limit for this system. The experimental resolution was determined by taking several knife edge cuts along the edges of the reconstruction and averaging the results. An experimental resolution of 128 ± 33 nm was obtained which is essentially the diffraction limit set by the zone plate. Interference fringes are visible beyond the numerical aperture of the zone plate. By estimating the effective numerical aperture from the interference pattern in Fig. 2(b), the resolution limit was calculated to be ~90 nm. This suggests that the zone plate focus is not centered within the pinhole and the pinhole scattered part of the reference wave to higher spatial frequencies, which in turn, improved the spatial resolution. Figure 2(d) shows a characteristic knife edge cut taken along the blue line in Fig. 2(c). The 10-90% intensity criterion was used to determine the resolution and is indicated by the black vertical lines in Fig. 2(d). It is apparent that the sample reconstruction was non-uniformly illuminated. This is produced by the multiple orders of the zone plate that generate an illumination wave with complex structure [28].

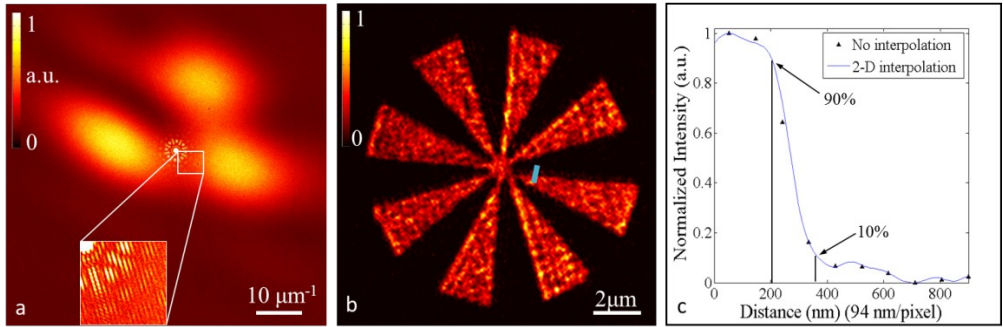

Fig. 3. (a) EUV hologram from a single ~1.2 ns pulse exposure. The inset shows a magnified view of the interference fringes. (b) Normalized reconstruction obtained by taking the modulus squared of the inverse Fourier transform of the hologram. (c) A characteristic knife edge cut used to calculate the overall resolution of the image. This knife edge cut was taken along the blue line in (b) indicating a resolution of 167 nm. The blue line was obtained by zero padding the hologram. The discrete points were obtained without any zero padding of the hologram.

The high flux of the tabletop EUV laser allowed us to demonstrate single-shot imaging. The experimental setup was similar to the previous section except the CCD was set to capture only a single laser pulse. A single-shot hologram taken from a ~1.2 ns laser pulse is shown in Fig. 3(a). The non-uniform illumination in Figs. 2(b) and 3(a) were due to the partial scattering of the reference wave off the pinhole. The pointing instability of the laser changes the focal spot location within the pinhole from shot to shot. The hologram in Fig. 2(b) was the integration of 20-shots with varying illumination patterns that resulted in a more complete and smooth illumination compared to Fig. 3(a). The lack of centrosymmetry in the holograms is a result of using a zone plate to create the reference. Friedel's law does not apply in this case due to the zone plate which introduces an additional phase into the reference wave. However, highly centrosymmetric diffraction patterns were observed when the zone plate was removed. The hologram shows high fringe visibility [Fig. 3(a). inset] and interference fringes extending almost to the edges of the CCD. The reconstruction is shown in Fig. 3(b). and a characteristic knife edge cut is shown in Fig. 3(c). The knife edge cut location is indicated by the blue line in Fig. 3(b). The resolution of 169 ± 58 nm was obtained by averaging several knife edge cut resolutions throughout the reconstruction. Typically, mechanical stability between the sample and zone plate would play an important role in determining the resolution limits, but an integration time of only 1.2 ns helped avoid this issue.

## 4. Conclusion

In conclusion, a coherent EUV imaging technique has been demonstrated that is capable of single-shot nanoscale resolution of an extended object. A multi-shot resolution of 128 ± 33 nm and a single-shot resolution of 169 ± 58 nm were obtained of an object ~11x11μm$^2$ in size. The resolution could be further improved by using a zone plate with a larger numerical aperture or by using a laser beam with a shorter wavelength. To improve the resolution beyond zone plate limitations, the method could be used in conjunction with a mask-based FTH or HERALDO. This would allow for larger objects to be imaged while maintaining the resolution limited by the reference features of the mask.

**Acknowledgments**

The authors acknowledge support by the Defense Threat Reduction Agency – Joint Science and Technology office for Chemical Biological Defense (Grant No. HDTRA1-10-1-007) and the National Science Foundation engineering research center for extreme ultraviolet science and technology award EEC 0310717.